# Gene expression rearrangements denoting changes in the biological state

Augusto Gonzalez (1,2), Joan Nieves (3), Maria Luisa Bringas Vega (1,4) and Pedro Valdes Sosa (1,4)

(1) The Clinical Hospital of Chengdu Brain Science Institute, MOE Key Lab for Neuroinformation, University of Electronic Science and Technology of China. Chengdu, China
(2) Institute of Cybernetics, Mathematics and Physics, Havana, Cuba
(3) Faculty of Physics, Havana University, Cuba
(4) Cuban Neurosciences Center, Havana, Cuba

**Abstract**
In many situations, the gene expression signature is a unique marker of the biological state. We study the modification of the gene expression distribution function when the biological state of a system experiences a change. This change may be the result of a selective pressure, as in the Long Term Evolution Experiment with E. Coli populations, or the progression to Alzheimer disease in aged brains, or the progression from a normal tissue to the cancer state. The first two cases seem to belong to a class of transitions, where the initial and final states are relatively close to each other, and the distribution function for the differential expressions is short ranged, with a tail of only a few dozens of strongly varying genes. In the latter case, cancer, the initial and final states are far apart and separated by a low-fitness barrier. The distribution function shows a very heavy tail, with thousands of silenced and over-expressed genes. We characterize the biological states by means of their principal component representations, and the expression distribution functions by their maximal and minimal differential expression values and the exponents of the Pareto laws describing the tails.

**Gene expression markers and distribution functions.** Present day technologies allow to measure gene expression (GE) levels in individual cells [1]. By means of techniques of dimensional reduction, such as principal component analysis (PCA [2,3,4]), one can show that the GE signature is a good marker of the cell state. Different options for the cell fate, for example, are seen to be resolved as disjoint regions in GE space [5].

With regard to tissues, although conceptually more complex because GE measurements in a small sample contain many different contributions, the procedure has proven its value, for example, in establishing spatial maps of the brain [6], in order to discriminate between a normal tissue and a tumor [7], etc. We believe that the GE signature could also be a good marker for the microstate of a tissue portion or sample, which takes account of the different cells entering the sample and the complex signaling system regulating the microenvironment.

In our paper, GE data is analyzed. The data comes from a long-term evolution experiment (LTEE) with E. Coli cultures [8], the Allen Institute study of aging and dementia [9], and The Cancer Genome Atlas (TCGA) [10]. In all of these experiments there are two well defined conditions: an initial or normal state, and a final or disease state. The evolution from initial to final states is precisely defined in the controlled experiment with bacteria. In the other two cases, however, the progression is not documented. Data from normal and disease samples is available and for their analysis we should assume a kind of ergodic hypothesis [11], stating that the microstates surveyed by the time evolution of a single sample as time becomes large enough coincide with those measured from many different samples at a given time.

We use PCA in order to characterize the systems states in GE space. In addition, we study how the GE distribution function is rearranged as the biological state changes from the initial or normal to the final or disease one.

Gene transcription, as any process in a living organism, is a noisy process [12], in which many small elements participate. The general result is a GE distribution function with a heavy tail [13], of power-like (Pareto) form [14].

We study how the expressions are redistributed as the biological state changes from initial to final or normal to disease. By geometric averaging over initial samples in order to smooth down the noise, we compute reference values for each gene in the initial state, $e_{ref}$. Differential expressions are defined as $d = e / e_{ref}$, and the distribution function in the final state, i.e. the number of genes with a given differential expression, is computed. $d \approx 1$ means that the expression level of a gene has not changed, whereas $d >> 1$ or $d << 1$ correspond to over-expressed or under-expressed (silenced) genes, respectively.

In the studied examples, we found two kinds of GE rearrangements after a change in the biological state. In the first case, most genes take values near the reference ones, and only a small fraction of genes take significant differential expression values. The distribution function is rapidly decaying as $d$ departs from 1. Because of the Pareto character, the decay law is $1 / d^{\nu}$, with a relatively large value for the exponent. This situation corresponds to relatively close initial and final states, and a ``continuous" transition.

The second general case, on the other hand, is characterized by radical expression rearrangements and heavy tails in the distribution functions (small exponents), involving thousands of differentially expressed genes. It corresponds to initial and final states far apart in GE space, and a ``discontinuous" transition.

In the next section, we use an analogy with physics in order to build up an intuition with regard to these two kinds of transitions.

**Continuous and discontinuous transitions in Physics.** In the upper left panel of Fig. 1, we draw a nearly harmonic potential well (dashed line). Under the action of a small amplitude noise, the motion of a particle in the well is characterized by a mean value for its position $<x> = 0$, corresponding to the potential minimum. This abstract picture may represent a biological system. The *x*-axis is a coordinate in GE space, and the *y*-axis is the fitness with a minus sign, such that the minimum of the potential is the state with maximal fitness.
INSERT FIGURE 1 AROUND HERE

Now, a small amplitude electric field is applied in the *x* direction. The resulting effective potential is drawn with a continuous line. A non zero minimum emerges. As time evolves, the result of the noisy motion is a mean position displacement from $<x> = 0$ to the new potential minimum. In the biological analogy, the electric field may be interpreted as a change in the external conditions, exerting new selection pressures. In the LTEE, for example, a fixed daily quantity of nutrients induce adaptation to this new conditions and a rise of fitness. The random noisy motion can be viewed as the result of mutations or epigenetic changes.

If the electric field is relatively small, the initial and final potential minima are relatively close to each other and the cloud described by the particle motion realizes a continuous transition between the minima (bottom left panel).

The second situation is depicted in the top right panel of Fig. 1. A double well with two distant minima is represented. The right minimum is deeper (higher fitness). This situation seems to describe cancer.

The initial (normal) state is prepared in the left well. It means that the particle starts realizing random motions from $\langle x \rangle = 0$. If the motions are of small amplitudes, the particle will remain in the left well for a long time because of the barrier preventing the transitions to the right well. Once a jump over the barrier takes place, the transition to a non zero mean value of $x$ occurs. It is seen as a discontinuous transition, or a jump in the mean position of the cloud described by the random particle motions (bottom right panel).

**Gene expression rearrangements in the LTEE.** The LTEE [8] is a formidable controlled evolution experiment with 12 E. Coli populations, followed for more than 60000 bacterial generations. We have studied some of the results coming from it [15,16] with the purpose of creating a model of mutations [17]. In the present section, we use the reported GE data [18], involving measurements in 4290 genes, in order to analyze the transition from the initial (ancestral) state to a final state at generation 20000. Data is provided for 8 harvested clones, coming from populations Ara+1 and Ara-1 in the experiment. 8 samples from the ancestral populations are also measured.

The conditions stressing the bacterial populations, i.e. the scarcity of nutrients, act since the very beginning of the experiment. The transition to the new state seems to be continuous, as suggested by the observed quasi-continuous variation of fitness as a function of time [19]. We shall verify how this transition is reflected in the PC representation and in the rearrangement of the GE distribution function.

We show the results of the PC analysis in the top panel of Fig. 2. A brief description of the procedures is given in the Methods section. We define new variables, $y = \log_2(d)$, from which the covariance matrix is constructed. Diagonalization of the matrix leads to new coordinate axes.
INSERT FIGURE 2 AROUND HERE

The PC1 axis, responsible for 43 % of the total data variance, seems to distinguish between the ancestral and evolved states. The coordinate $x_1$ is the projection along PC1, that is $x_1 = \mathbf{y} \cdot \boldsymbol{u_1}$, where $\boldsymbol{u_1}$ is the normalized vector along the PC1 axis.

The mean value of the $x_1$ coordinate changes from $\langle x_1 \rangle = 0$ to $\langle x_1 \rangle = 21.44$. The mean radii of the ancestral and evolved clouds of samples, measured from the standard deviations along the PC1 axis, are 3.08 and 12.77, respectively.

Let us stress that the evolved state at generation 20000 may be seen as an intermediate stage in the transit between minima in Fig. 1. Indeed, the fitness keeps increasing at least until generation 50000 [19].

The central and bottom panels show the GE distribution functions. They are integrated distribution functions, that is count the number of genes with differential expression greater (lower) than a given value. Notice that the slope of the over-expression log-log curve for $d < 2$ (the Pareto exponent) is around -10, whereas the slope in the under-expression curve for $d > 1/3$ is around 4. At these points, there are changes in the exponents to values -2 and +2, respectively (the dotted lines).

There are only 4 genes in the extreme region $d > 2$ (in the Ara+1 culture), and around 20 genes in the opposite region $d < 1/3$. The total number of differentially expressed genes should be contrasted with the around 30 beneficial mutations detected at generation 20000 [15,16]. Up to this point, gain of fitness is achieved mainly by turning off non active metabolic processes, i.e. by silencing the responsible genes [18].

Summarizing the section, we may say that in the experimentally observed continuous transition in the LTEE, the initial (ancestral) and the final (evolved) states are relatively close in GE space, and the GE distribution functions of both states are also close, with only around 25 genes exhibiting significant values for the differential expression, that is a fraction of around 1/200 of the total number of genes. The latter criteria will be employed to assess the continuous character of the transition in the example studied in the next section.

**Changes in brain white matter and Alzheimer disease.** The second studied example is the GE data obtained post-mortem from a cohort of patients with Alzheimer disease (AD) and non-demented controls (ND), whose ages are above 77 years. The data comes from the Aging, Dementia and TBI study by the Allen Institute [9,20]. http://aging.brain-map.org/download/index

In the Allen study, samples are collected from four brain regions known to show neurodegeneration and be related to pathologies as a result of AD and Lewis body disease (as described in [20]): temporal and parietal neocortex (TCx and PCx), hippocampus (HIP) and white matter of the forebrain (FWM).

A general PCA picture of AD and ND samples is drawn in Fig. 3. It is apparent that in the neocortex and the hippocampus, the clouds of ND and AD samples practically overlap. Samples from the white matter, on the other hand, are distributed over a wider sector in GE space, and it seems to be a clear distinction between the AD and ND zones.
INSERT FIGURE 3 AROUND HERE

Thus, below we focus on FWM. There are 47 ND and 28 AD samples, coming from different patients. The number of involved genes in the study is 50281.

The upper panel of Fig. 4 shows the results of the PC analysis of the FWM data.
INSERT FIGURE 4 AROUND HERE

The PC1 axis, which accounts for 24.7 % of the total data variance, discriminates between the ND and AD states. The transition between both states is accompanied by a change from $\langle x_1 \rangle = 0$ to $\langle x_1 \rangle = 40.97$. However, the radii of the ND and AD clouds of samples are larger than the inter-center distance, that is 80.69 and 72.64, respectively. These results suggest a continuous transition in a very broad well.

It is well known the role of age in AD, specially in the elderly [21]. Then, we may use age as a time variable to follow the transition. In spite of the relatively small number of samples, a linear regression analysis of the mean position $\langle x_1 \rangle$ as a function of age in ND samples, Fig. 5 top panel, shows that $\langle x_1 \rangle \approx -287.12 + 3.24\, age$, p value = 0.07. In the AD samples, however, no correlation between $\langle x_1 \rangle$ and *age* is observed. Thus, the position of the AD zone is roughly fixed, and the cloud of ND samples shows a drift towards the AD minimum as age increases.

INSERT FIGURE 5 AROUND HERE

A better illustration of this fact comes from Fig. 6, where the probability density of ND and AD samples along the PC1 axis is compared. Four age intervals, containing roughly the same number of ND samples are defined: [77,84], [84,90], [90,95] and [95,100+] years. The total AD probability is shown in the four panels. A drift towards the AD zone as a result of aging is apparent.
INSERT FIGURE 6 AROUND HERE

The bottom panel of Fig. 5 shows the increase with age for both ND and AD samples of the NIA Reagan index for the neuropathological diagnosis of AD [22]. This may be simply interpreted as an increase of the fraction of brain microstates trapped in the AD zone.

Let us recall the physical analogy, mentioned above. The random motion of samples in GE space can not be ascribed to mutations because it is well known that the replacement rate of neurons is very low [23]. These random displacements or variations in GE space are instead related to accumulation of damage in the DNA of brain cells [24] or to accumulation of methylation events [25,26], both processes are related to aging and in general lead to a decrease of tissue fitness. The roughly independent of age position of the AD zone means that this is a definite region in GE space with higher fitness, a local maximum, which holds the disease state.

The following picture of late AD progression emerges. As age increases, the fitness of brain micro-regions decrease and a zone of GE space representing a local maximum (the AD zone) becomes reachable. The neocortex and other brain regions are attracted earlier to this zone. The white matter, responsible for the connections and probably defining the global AD brain state, shows higher resilience. Below, we should come back to this picture.

The central and bottom panels of Fig. 4 illustrate the rearrangement in the expression levels. The distribution function exhibits a fast decay when the differential expression departs from 1. The exponents of the Pareto laws are -8 and 9, respectively. There are around 100 genes with $d > 2$, and only around 10 genes with $d < 1/2$. The fraction of differentially expressed genes is ~ 1/500.

Summarizing the section, we may say that the data on GE in the white matter of aged brains seems to support a picture of a continuous transition from the ND to the AD state motivated by a modification of the potential (the fitness distribution) at ages below 77.

**The transition from a normal tissue to a tumor.** In this section, we consider a set of human tissues. In a lifetime span, the stem cells of some of them realize around 10000 divisions [27,28]. If the tissue is in a tumor phase, an increase of the division rate is expected [29]. Thus, with respect to the number of cell divisions (generations), the data for tumor cells are comparable to that of bacteria.

We analyze GE data from the TCGA [10] for the 15 tumor localizations described in Table 1. Expression levels for 60483 genes are measured. Normal and tumor samples from different patients are recorded. Thus, we should make use of the ergodic hypothesis for the analysis of the data. We stress that a set of results coming from the PCA of this data is presented in Ref. [30]. Below, we focus on the rearrangements of GE levels.

Let us consider the Kidney Clear Cell Carcinoma (KIRC) in more details. The PC analysis is presented in the top panel of Fig. 7. The PC1 axis, responsible for 60 \% of the data variance, discriminates between normal and tumor samples. The mean value of the $x_1$ coordinate varies from $\langle x_1 \rangle = 0$ to $\langle x_1 \rangle = 171.80$ in the transition from the normal to the tumor state. The radii of these regions are 28.70 and 36.00, respectively. Thus, the data suggests that there exist two distinct minima, occupying distant regions in GE space.
INSERT FIGURE 7 AROUND HERE

According to the ergodic hypothesis, the higher density of observed samples correspond to the potential wells (higher fitness regions). The deepest well seems to be the tumor state. The intermediate region $30 < x_1 < 130$, supports a low-fitness barrier which prevents the transition from the normal to the tumor state. In particular, $30 < x_1 < 80$ defines a coexistence region, where both normal and tumor samples are observed.

In our previous paper [31], we have quantitatively estimated the number of available microstates in each region for a set of tumors by means of an entropy-like magnitude.

The progression of a normal sample to a tumor state could proceed as follows. The sample starts at a point near $x_1 = 0$ and realizes random motions due to somatic mutations, epigenetic changes or external carcinogenic factors. However, the barrier prevent the sample from leaving the normal region. Only when a jump over the barrier occurs the sample starts moving towards the tumor region.

In the central and bottom panels of Fig. 7, we show the distribution function for the differential expressions in the over- and under-expression regions. The average tumor curves exhibit exponents near -1.4 and 0.7, respectively, and there are thousands of differentially expressed genes. These results support the picture of a discontinuous normal tissue to tumor transition.

Two additional curves were added to this figure. They reflect the average distributions of normal and tumor samples in the intermediate coexistence region, and show how the rearrangement of expression levels occurs in the progression to tumors. The greatest differences between normal and tumor distributions become apparent in the under-expression region. Roughly speaking, these are genes related to homeostasis, which are silenced in the tumor state.

The results for the other tumor localizations, studied in the present paper, are summarized in Table 1. The mean value of the $x_1$ coordinate in the tumor state (for the normal state we set $\langle x_1 \rangle = 0$), the radii of the normal and tumor zones, the Pareto exponents, and the maximal and minimal reached differential expression values are given for each tissue.

We have grouped in a final Fig. 8 the distribution functions for all of the studied tumor localizations, which shows a kind of universal behavior in cancer.
INSERT FIGURE 8 AROUND HERE

Summarizing the section, we may say that the transition from a normal tissue to a tumor seems to be a discontinuous one. The differential distribution functions show very heavy tails with thousands of differentially expressed genes, around 1/10 of the total number of genes.

**Concluding remarks.** We use an analogy with the motion of a particle realizing random displacements in an external potential in order to analyze GE rearrangements in a biological system which experiences a transition from an initial to a final state. The random motion of the particle is associated to variations in the expressions of a group of genes as a result of mutations and epigenetic events, or even damages in the DNA. The external potential is the fitness landscape.

In the LTEE, the experiment conditions induce displacement towards a new minimum, away from the initial one corresponding to the wild or ancestral genotype.

In the study concerning late onset of AD, we observe an AD zone with a definite position in GE space, and a drift of the ND clouds of samples towards the AD zone as age increases.

Both are examples of continuous transitions, motivated by a modification of the fitness landscape. This modification is well understood in the LTEE. In the AD study, on the other hand, we think that the accumulation of damages and methylation events as a result of aging is not only the reason for the random motion in GE space, but leads also to a significant reduction of fitness in the microstates. Recalling the fitness landscape in the next example, tumors, we may say that aging makes the brain microstates to move away from the normal, homeostatic zone to the low-fitness region. It seems that the AD zone is located somewhere in this region and is a kind of local maximum for the fitness, to which the ND samples are attracted.

The idea of aging as a cause for reaching the low-fitness barrier is also consistent with the increase of cancer risk with age.

Continuous and discontinuous transitions are reflected in different ways in the GE distribution functions. The former corresponds to slight, whereas the latter corresponds to radical rearrangements.

We quantitatively describe the geometry of minima in GE space, and the tails of the GE distribution functions.

**Methods.** The GE data corresponding to the studied examples is analyzed by means of the PCA technique. The details of the PC analysis may be found in paper [30]. Me briefly sketch them in the present section.

The dimension of matrices in the Principal Component analysis is equal to the number of genes in the data. Usually, in order to compute the average expression of a gene the median or the geometric mean are used. We prefer geometric averages, but then the data should be slightly distorted to avoid zeroes. To this end, we added a constant to the expression (0.0001 in the LTEE data, 0.1 in the other two examples). By applying this regularization procedure, genes identified as relevant could be under question if the differential expression is relatively low and their expression in the normal or initial state is near zero. As we are mainly interested in the strongly over- or under-expressed genes, they are out of the question.

We take the mean geometric average over normal or initial state samples in order to define the reference expression for each gene, $e_{ref}$. Then the normalized or differential expression is defined as: $d = e / e_{ref}$. The fold variation is defined in terms of the logarithm $y = \log_2 (d)$. Besides reducing the variance, the logarithm allows treating over- and sub-expression in a symmetrical way.

Deviations and variances are measured with respect to $y = 0$. That is, with respect to the average over normal samples. This election is quite natural, because normal samples are the majority in a population.

With these assumptions, the covariance matrix is written:

$$\sigma_{ij} = \sum y_i(s)\, y_j(s) \;/ (N_{samples} - 1), \qquad (1)$$

where the sum runs over the samples, $s$, and $N_{samples}$ is the total number of samples (initial or normal plus final or disease). $y_i(s)$ is the fold variation of gene $i$ in sample $s$.

As mentioned, the dimension of matrix $\sigma_{ij}$ coincides with the number of genes in the data. By diagonalizing it, we get the axes of maximal variance: the Principal Components (PCs). They are sorted in descending order of their contribution to the variance.

In Lung Squamous Cell tumors (LUSC), for example, PC1 accounts for 67 % of the variance. This large number is partly due to our choice of the reference, $y = 0$, and the fact that most of the samples are tumors. The reward is that PC1 may be defined as the cancer axis. The projection over PC1, i.e. the coordinate $x_1 = \mathbf{y} \cdot \boldsymbol{u_1}$, where $\boldsymbol{u_1}$ is the unitary vector along PC1, defines whether a sample is classified as normal or tumor.

The next PCs account for a smaller fraction of the variance. PC2 is responsible of 4 %, PC3 of 3 %, etc. Around 20 PCs are enough for an approximate description of the region of the gene expression space occupied by the set of samples.

We want to compute only a small number of the eigenvalues and eigenvectors of $\sigma$. To this end, we use a Lanczos routine in Python language, and run it in a node with 2 processors, 12 cores and 64 GB of RAM memory. As a result, we get the first 100 eigenvalues and their corresponding eigenvectors.

**Acknowledgments.** A.G. acknowledges the Cuban Program for Basic Sciences, the Office of External Activities of the Abdus Salam Centre for Theoretical Physics, and the University of Electronic Science and Technology of China for support. The research is carried on under a project of the Platform for Bioinformatics of BioCubaFarma, Cuba.

**Funding:**
The participation of AG and PVS was partially funded by the International Academician Station for Global Precision Medicine and the Project number Y03111023901014005 at the University of the Electronic Sciences and Technology of China UESTC.

| Tissue | $<x_1>$ | $R_n$ | $R_t$ | $d_{min}$ | $\upsilon_{under}$ | $d_{max}$ | $\upsilon_{over}$ |
|---|---|---|---|---|---|---|---|
| BRCA | 137.37 | 20.97 | 31.66 | 0.0132 | 2 | 70.15 | -1.6 |
| COAD | 155.89 | 11.71 | 28.53 | 0.0032 | 1.6 | 60.41 | -2 |
| HNSC | 123.5 | 27.74 | 23.54 | 0.0087 | 1.5 | 45.15 | -2.5 |
| KIRC | 171.8 | 28.7 | 36 | 0.0002 | 0.7 | 299.6 | -1.4 |
| KIRP | 163.42 | 19.9 | 27.78 | 0.0001 | 0.8 | 43.16 | -2.5 |
| LIHC | 134.67 | 20.48 | 45.23 | 0.0113 | 1.8 | 40.34 | -3 |
| LUAD | 145.33 | 13.51 | 32.06 | 0.0034 | 1.8 | 41.71 | -2.8 |
| LUSC | 194.49 | 11.62 | 36.65 | 0.0009 | 1.7 | 364.5 | -1.5 |
| PRAD | 91.33 | 31.31 | 32.17 | 0.0439 | 3.2 | 21.81 | -3.2 |
| STAD | 136.97 | 27.14 | 43.24 | 0.0138 | 1.8 | 71.84 | -3.5 |
| THCA | 112.54 | 20.02 | 39.85 | 0.0154 | 2 | 62.52 | -2 |
| UCEC | 171.38 | 38.24 | 22.14 | 0.0054 | 1.7 | 80.45 | -2 |
| BLCA | 140.61 | 57.53 | 34.68 | 0.0061 | 1.8 | 18.28 | -3.5 |
| ESCA | 138.7 | 64.28 | 35.79 | 0.001 | 0.8 | 29.78 | -3 |
| READ | 168.05 | 22.9 | 28.81 | 0.0021 | 1.5 | 171.8 | -1.5 |

**Table 1.** The studied cancer localizations and the main results of the section. $<x_1>$ is the position along PC1 of the center of the cloud of tumor samples. $R_n$ and $R_t$ are the radii, measured from the standard deviations, of the normal and tumor clouds of samples, respectively. $d_{min}$ and $d_{max}$ are the minimal and maximal values of the differential expressions, and $\upsilon_{under}$ and $\upsilon_{over}$ the Pareto exponents in the under-and over-expression regions. TCGA abbreviations for the tumors are used.

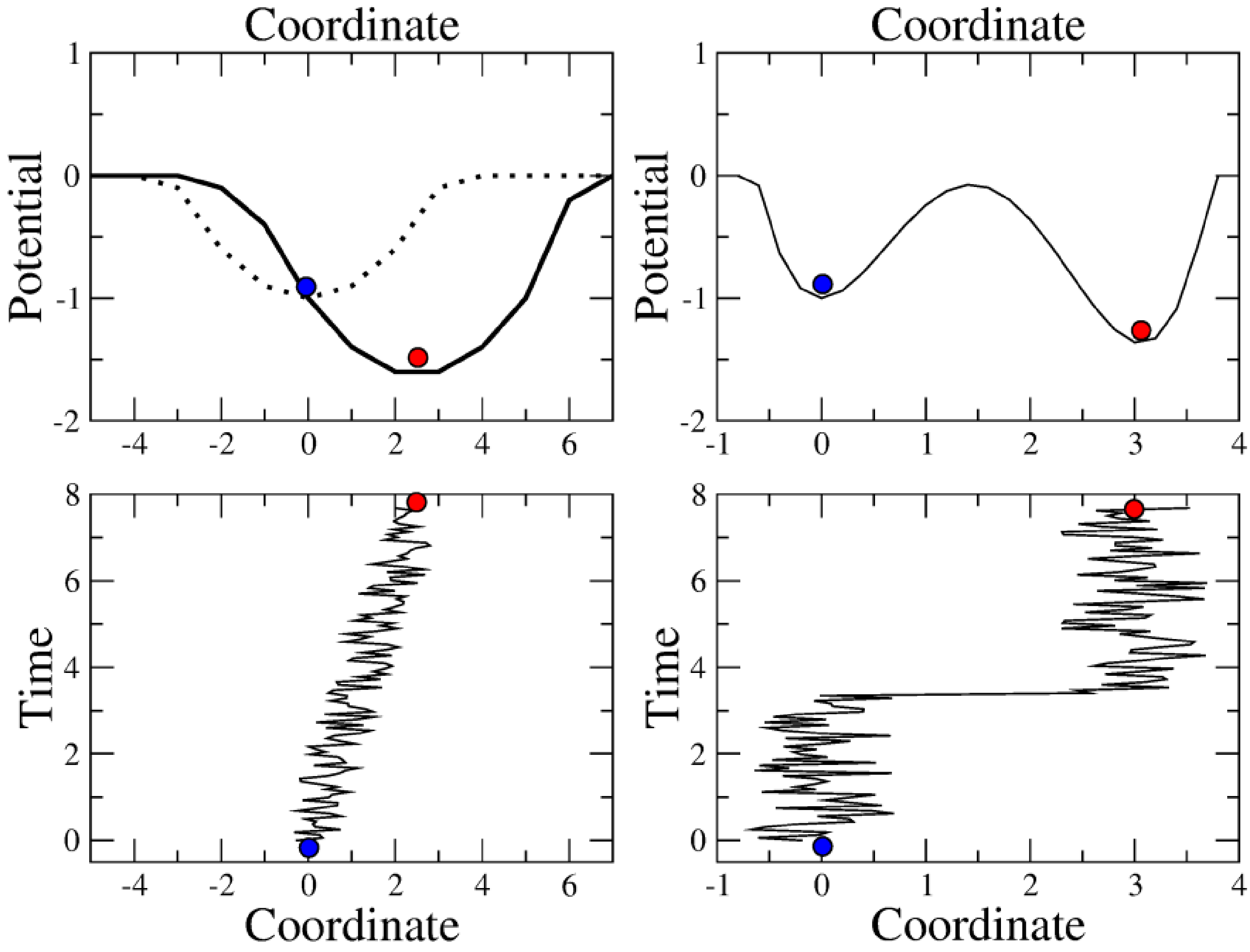

**Fig. 1.** Left: Illustration of a continuous transition. The addition of a small electric field to a harmonic potential well causes a modification of the minimum from $\langle x_1 \rangle = 0$ (blue circle) to a nonzero value (red circle). Right: Illustration of a discontinuous transition in a double well with distant minima. Random fluctuations may drive a particle, initially in the left well, towards the right deepest well. The barrier separating the two minima should be surpassed.

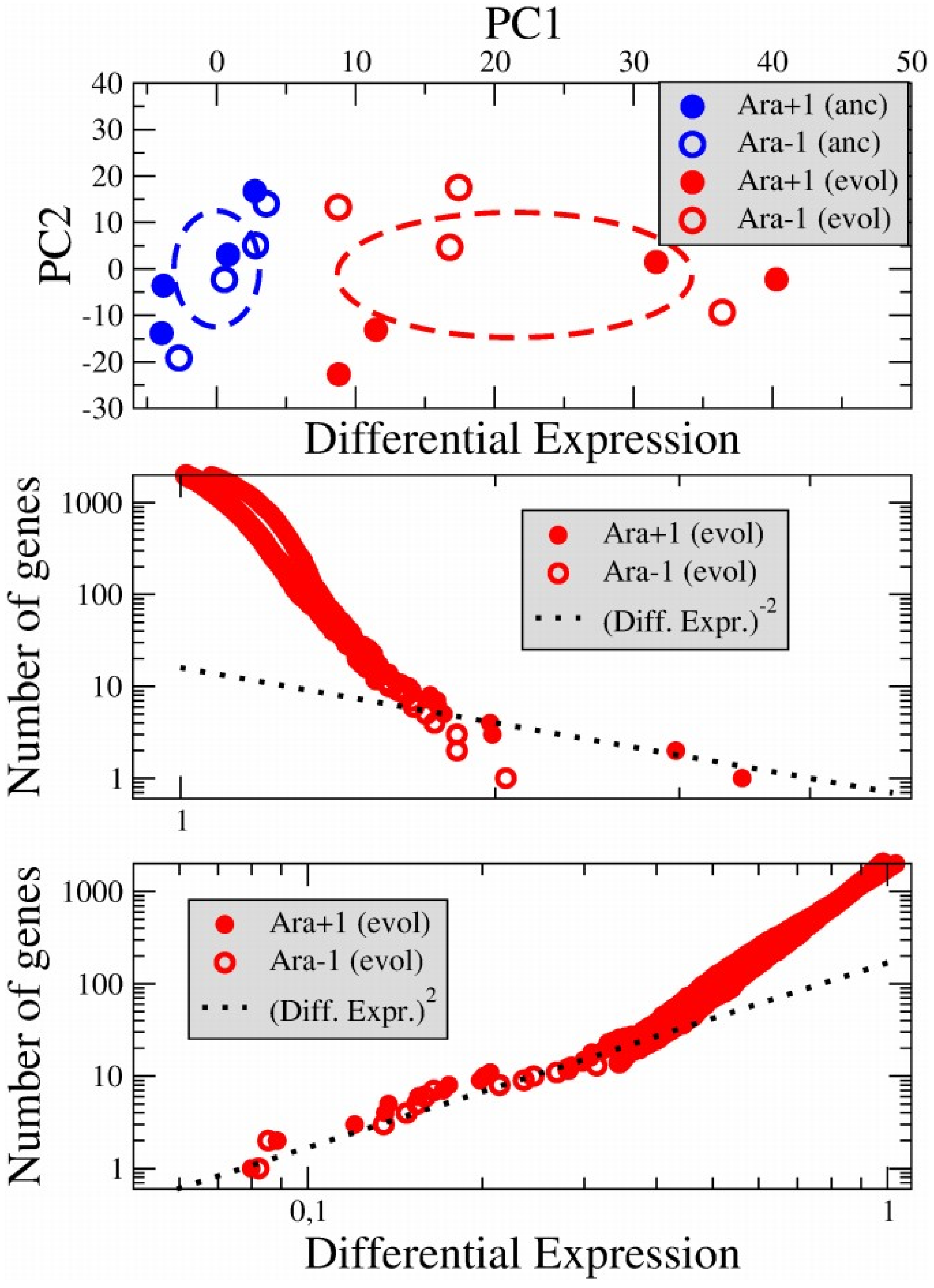

**Fig. 2.** Top: Principal component analysis of the gene expression data in the LTEE. Samples from the ancestral (blue circles) and evolved populations (at generation 20000, red circles) are shown. Dashed ellipses are drawn according to the standard deviations in each zone. Center and bottom: Rearrangement of the gene expression levels as a consequence of the evolution.

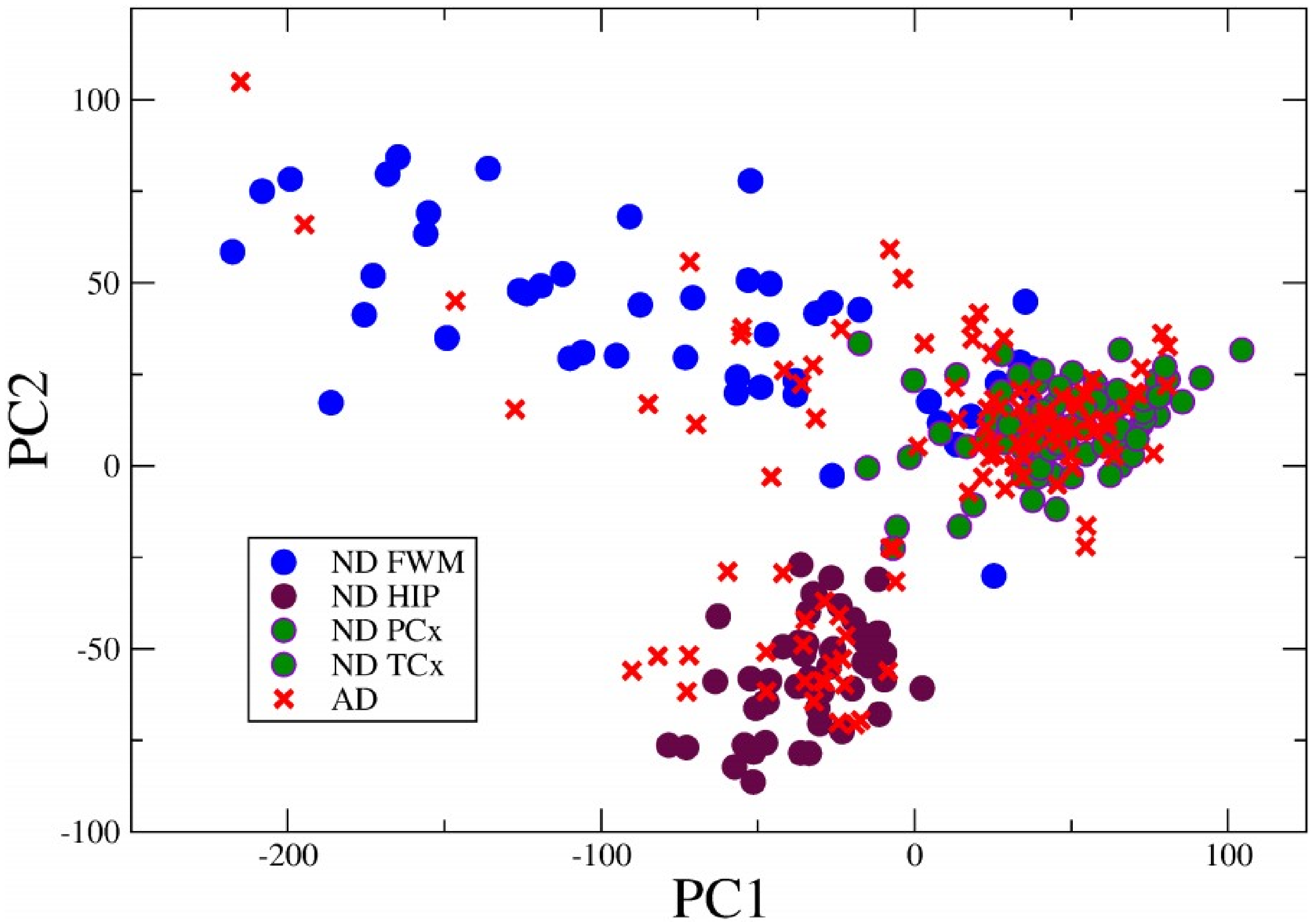

**Fig. 3.** Principal component analysis of the Allen Institute GE data on AD. ND and AD samples are shown in each of the four studied brain regions: FWM, HIP, PCx and TCx. The best separation between ND and AD states occurs in FWM, in the other three regions the clouds of ND and AD samples practically overlap.

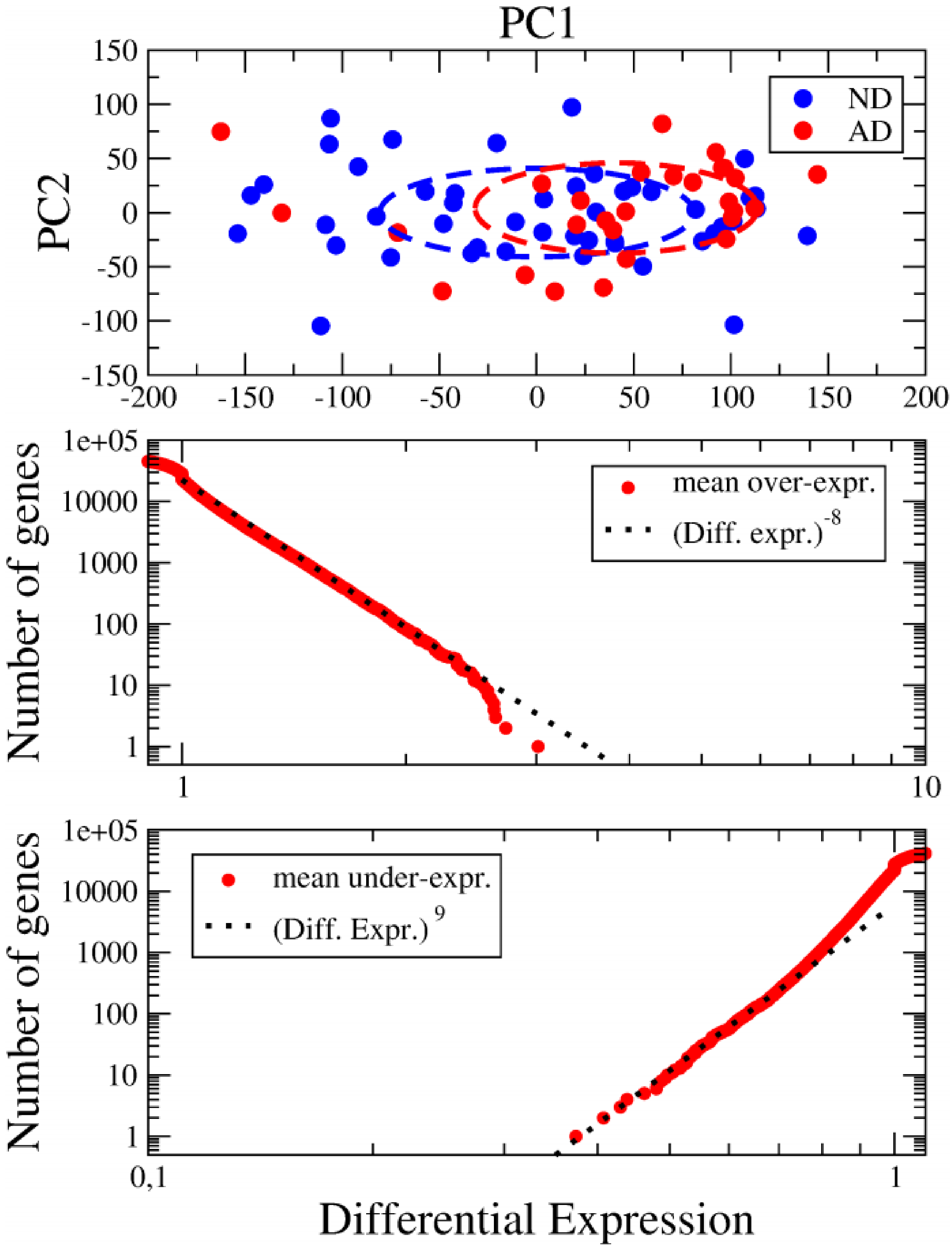

**Fig. 4.** Top: Principal component analysis of the Allen Institute gene expression data in FWM. ND and AD samples are shown. Dashed ellipses are drawn according to the standard deviations in each zone. Center and bottom: Differential gene expression distribution functions in the AD state. Only a few dozens of genes reach significant differential expression values.

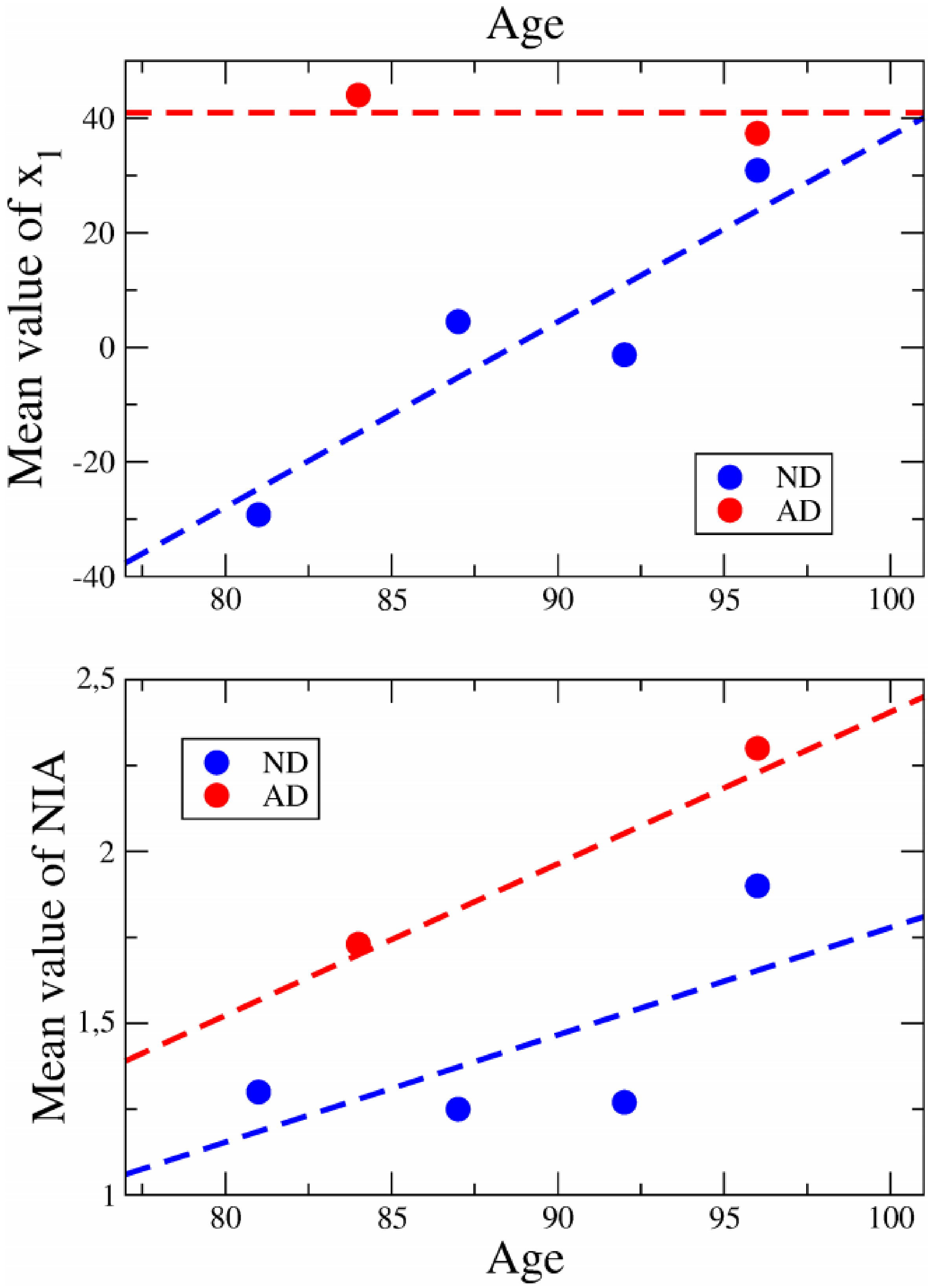

**Fig. 5.** Top: Mean sample position along the PC1 axis as a function of age. As age increases, the ND samples experience a drift towards the AD region, which center is roughly age independent. Botton: Age dependence of the NIA Reagan index in AD and ND samples. The BRAAK stage and CERAD score exhibit similar dependences on age.

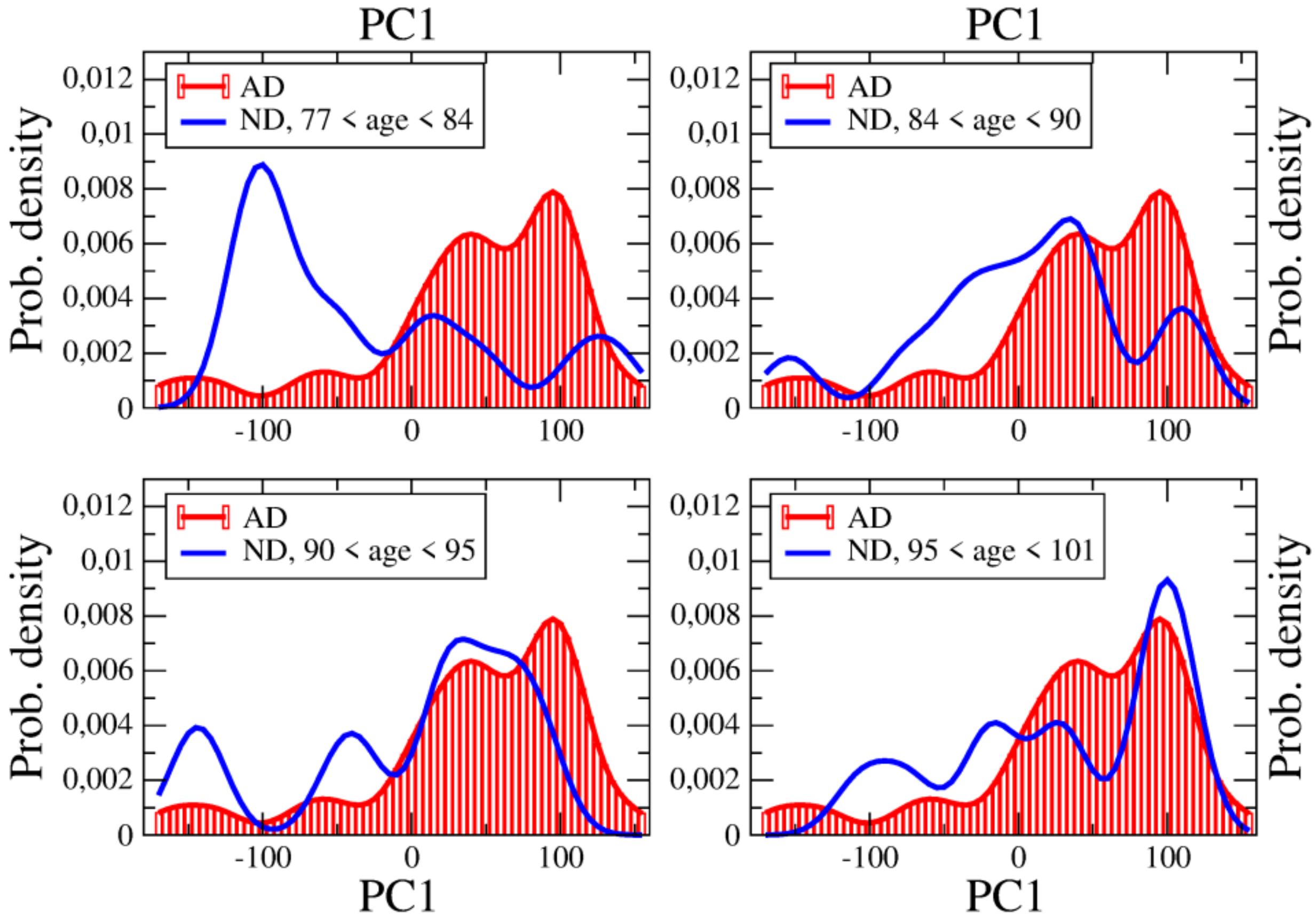

**Fig. 6.** Probability density of ND and AD samples along the PC1 axis. Each panel refers to an age interval for ND samples. The AD probability, which is roughly age independent, is shown in the four panels.

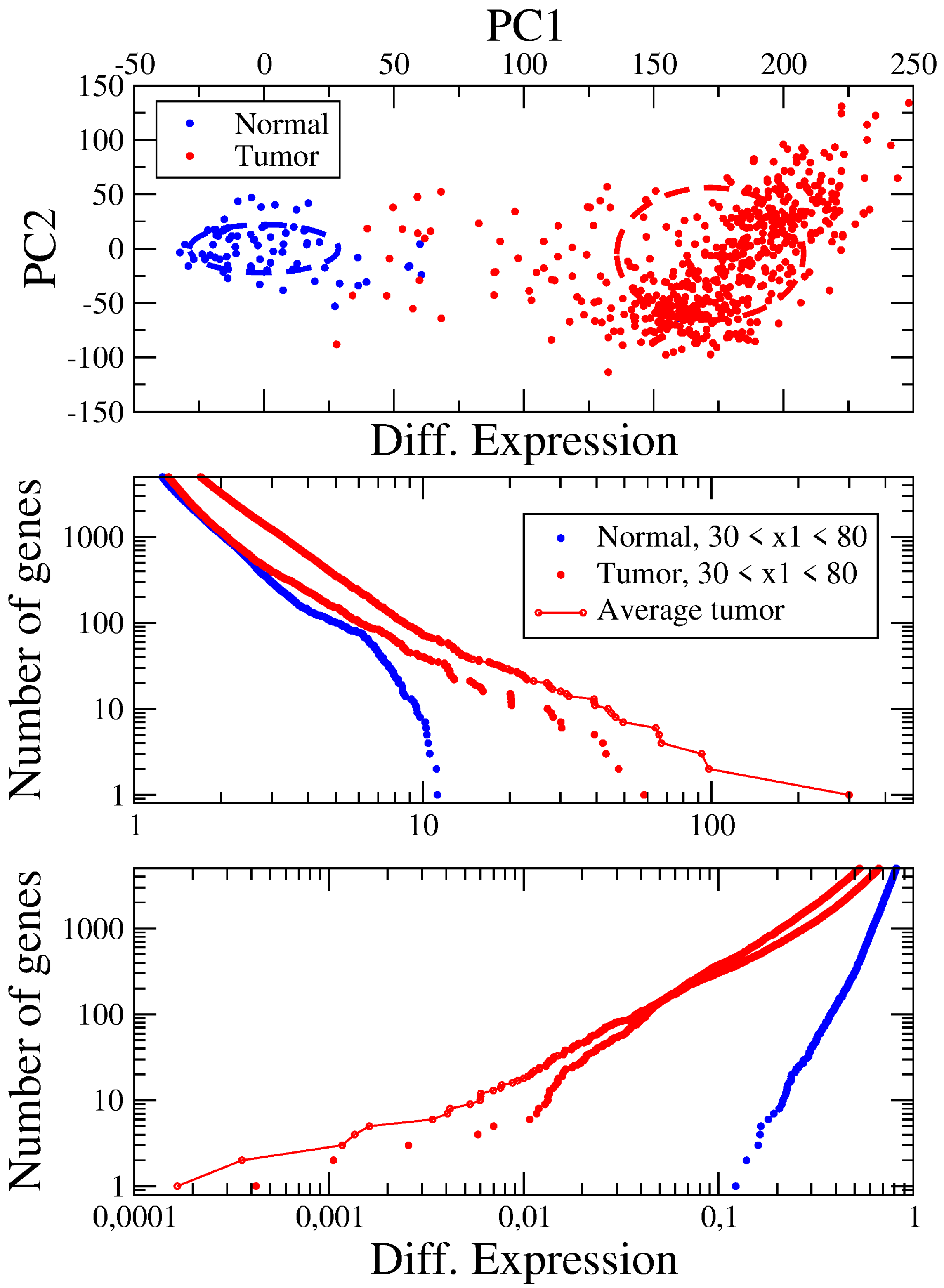

**Fig. 7.** Upper panel: Principal component analysis of the gene expression data in Clear Cell Kidney Cancer (KIRC). PC1 is the axis describing the progression from the normal to the tumor state. Dashed ellipses are drawn according to the standard deviations in each zone. Center and lower panels: Rearrangement of the gene expression distribution function in the progression from normal tissue to tumor.

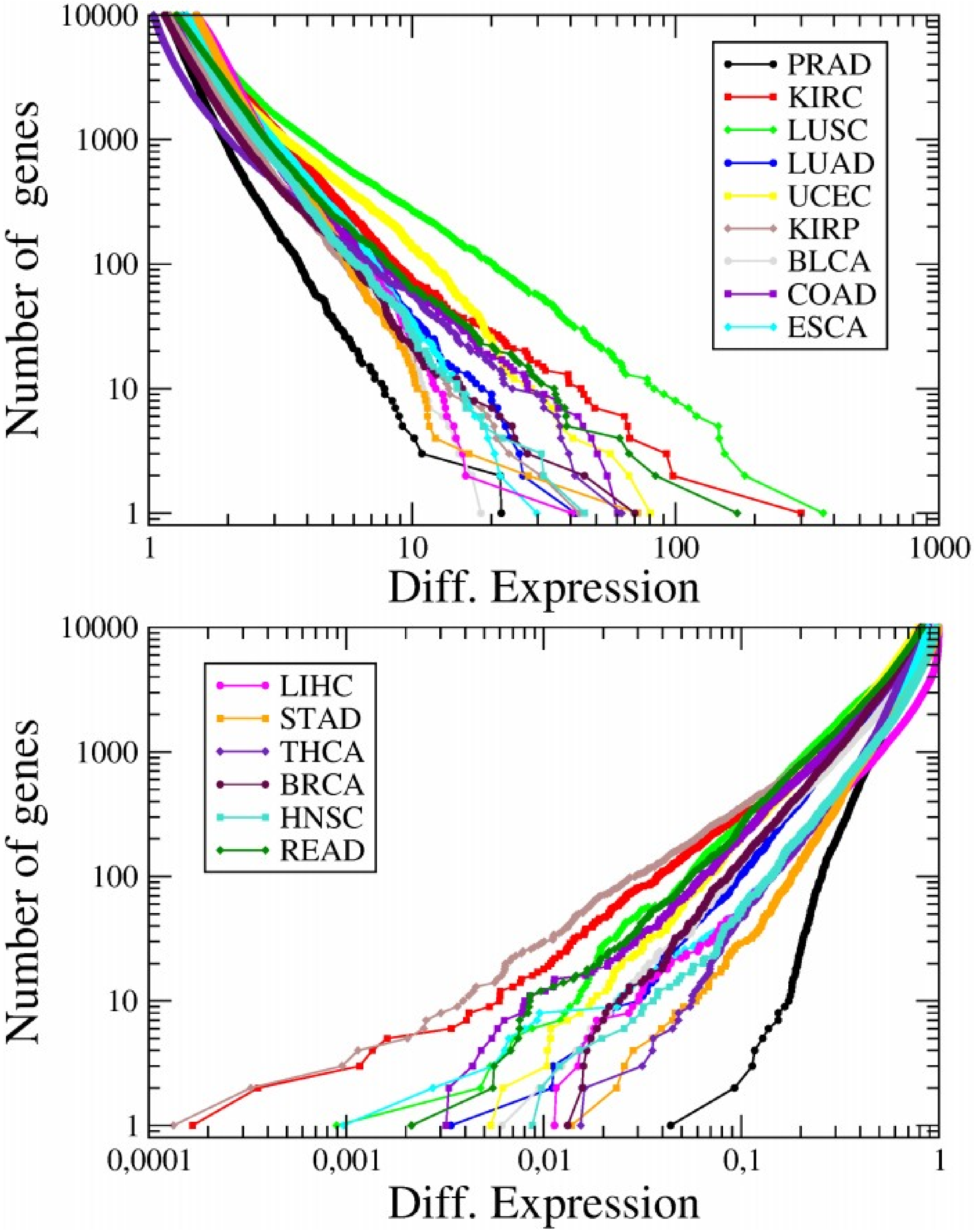

**Fig. 8.** The differential gene expression integrated distribution function for 15 tumor localizations in the TCGA. Pareto exponents between -1.4 and -3.5 in the over-expression tails, and between 0.7 and 2.0 in the under-expression tails are observed.